\documentclass[prc,twocolumn,showpacs,preprintnumbers,amsmath,amssymb,superscriptaddress,floatfix,nofootinbib]{revtex4}

\usepackage{graphicx}
\usepackage{amsmath}
\usepackage{amsfonts}
\usepackage{amssymb}%
\def\etal{{\em et al.}}

\begin{document}

\title{Evidence for a narrow $D_{03}$ state in $K^- p \rightarrow \eta \Lambda$ near threshold}

\author{Bo-Chao Liu} \email{liubc@xjtu.edu.cn} \affiliation{Department of Applied Physics, Xi'an
Jiaotong University, Xi'an, Shanxi 710049, China}
\affiliation{Theoretical Physics Center for Science Facilities,
Chinese Academy of Sciences, Beijing 100049, China}

\author{Ju-Jun Xie} \email{xiejujun@mail.ihep.ac.cn}
\affiliation{Department of Physics, Zhengzhou University, Zhengzhou,
Henan 450001, China} \affiliation{Theoretical Physics Center for
Science Facilities, Chinese Academy of Sciences, Beijing 100049,
China}

\begin{abstract}
Recently, we reported a theoretical study on the $K^- p \to \eta
\Lambda$ reaction near threshold by using an effective Lagrangian
approach. It was found that the description of angular distribution
data measured by the Crystal Ball Collaboration needs a $D_{03}$
resonance with mass $M=1668.5\pm 0.5$ MeV and total decay width
$\Gamma=1.5\pm 0.5$ MeV, which is not the conventional
$\Lambda$(1690) or other $\Lambda$ state listed in the Particle Data
Group book. In the present work, we study the $\Lambda$ polarization
in the $K^- p \to \eta \Lambda$ reaction within the same framework.
The results show that the existence of this narrow $D_{03}$ state is
also compatible with current $\Lambda$ polarization data and that the
more accurate $\Lambda$ polarization data at $P_{K^-}=735$MeV can
offer further evidence for the existence of this resonance.
Furthermore, the role of the $\Lambda(1690)$ resonance in this reaction
is also discussed.

\end{abstract}
\maketitle

\section{Introduction}

Understanding $\bar KN$ interactions in the low-energy region
is a very important part of the study of the behavior of quantum
chromodynamics(QCD) in the non-perturbative regime. Because the
excitation of hyperon resonances usually dominates in relevant
processes, it offers a good basis to study the properties of hyperon
resonances, especially for hyperon resonance with $S=-1$. In fact,
up to now most of the knowledge of hyperon resonances is from the
analysis of $\bar KN$ interactions~\cite{pdg2012}. Even though we have
studied the $\bar KN$ interactions for a long time, the large
uncertainties of the parameters of hyperon resonances in the Particle
Data Group~(PDG) book indicate that the status of our knowledge on
these resonances is still not satisfying. This is partly because of
the complications of the non-perturbative character of QCD in the low-energy regime and partly because of the poor quality of the
available experimental data.

In the past ten years, some new experimental data on $K^-p$
scattering with much higher accuracy than before were reported by the
Crystal Ball Collaboration~\cite{data,Klempt2010}. With these new
data, from theoretical analysis, it is possible to refine our
knowledge on the hyperon resonances and to better understand the
mechanism of relevant reactions. Some work along this way has
already been done\cite{gaopz2011,lam1670prl,oset1,oset2,zhongprc79,Liu}.
Among various $\bar KN$ inelastic reactions, the reaction $K^- p
\rightarrow \eta \Lambda$ is particularly interesting and important.
Due to isospin conversation, the $\Sigma$ resonances do not
contribute in this reaction, which makes this reaction a good place
to study the properties of $\Lambda$ resonances. On the other hand,
our current knowledge of the couplings of $\Lambda$ resonances with
the $\eta\Lambda$ channel is still very poor. Besides the
$\Lambda(1670)$ resonance, the couplings of other $\Lambda$
resonances with the $\eta\Lambda$ channel are only poorly known or
unknown. Some further studies on this reaction are obviously still
needed.

In our previous work~\cite{Liu}, we analyzed the new data reported
by the Crystal Ball Collaboration for the $K^- p \to \eta \Lambda$
reaction~\cite{data} within an effective Lagrangian approach. It is
found that by including the background and $\Lambda(1670)$
resonance the total cross section data can be well reproduced.
However, the bowl structure appearing in the differential cross
section data cannot be explained. We showed that the differential
cross section data favor a $D_{03}$ resonance with very narrow
width, which is not the conventional $\Lambda(1690)$ resonance or
other $\Lambda$ states listed by the PDG~\cite{pdg2012}. As mentioned in
Ref.~\cite{Liu}, the current experimental data still have systematic
uncertainties, especially when we look at the angular distribution
data obtained from two different ways of identifying the final
$\eta$ meson (see Fig.20 of Ref.~\cite{data}). For better
understanding of the origin of the higher partial wave contributions,
we suggest our experimental colleagues remeasure the angular
distribution data.

On the theoretical side, it is also important to find some other way
or criterion to check for the existence of this narrow $D_{03}$
resonance. In the present work, we give predictions of the $\Lambda$
polarization in the $K^- p \to \eta \Lambda$ reaction by using the
parameters determined from the fitting to differential cross section
data in our previous work~\cite{Liu}. On the other hand, since the
conventional $\Lambda(1690)$ resonance may also contribute to this
reaction in principle, we also give some further discussion about
the role of the $\Lambda(1690)$ resonance in this reaction. The present
work can be treated as a further test of our previous
results~\cite{Liu} and offer some new criterion to verify the
existence of this narrow $D_{03}$ resonance.

The paper is organized as follows. In the next section, we present the
ingredients and formalism used for the present calculation, while
the results with some discussion are given in Sec.~III. Then, a
short summary is given in the last section.

\section{Ingredients and formalism}

\begin{figure}[htbp] \vspace{0.cm}
\begin{center}
\includegraphics[scale=0.5]{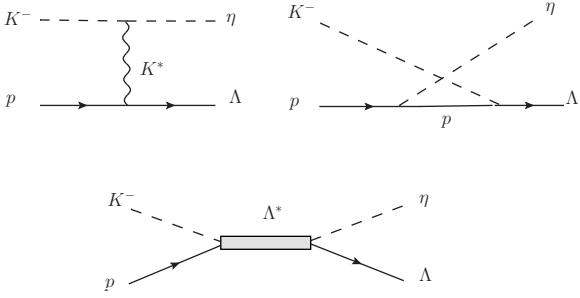}
\caption{Feynman diagrams for the reaction $K^- p \to \eta \Lambda$.
The t-channel $K^*$ meson exchange, u-channel proton exchange, and
the s-channel $\Lambda$ resonance exchange are considered.}
\label{feynfig}
\end{center}
\end{figure}

We adopt the effective Lagrangian method in describing the reaction
$K^- p \to \eta \Lambda$ near threshold. The effective Lagrangian
method is an important theoretical approach in studying various
processes in the resonance region, and it is widely used in partial
wave analysis for the properties of resonances. The main
contributions for this reaction come from the t-channel $K^*$ meson
exchange, the u-channel proton exchange, and the s-channel $\Lambda$
resonance exchange. The corresponding Feynman diagrams are shown in
Fig.~\ref{feynfig}.

First, for the t-channel $K^*$ meson exchange, we take the
normally used effective Lagrangians for $K^*K\eta$ and $K^*N\Lambda$
couplings as

\begin{eqnarray}
{\cal L}_{K^* \bar K \eta} &=&
g_{K^*K\eta}(\eta\partial^\mu{K^-}-K^-\partial^\mu{\eta})K^{*-}_\mu.
\label{ksnl} \\
{\cal L}_{K^*N\Lambda} &=& g_{K^*N\Lambda}\bar\Lambda\big(\gamma_\mu
K^{*\mu}-{\kappa_{K^*N\Lambda}\over 2M_N}\sigma_{\mu\nu}\partial^\nu
K^{*\mu}\big)N  \nonumber \\
&& + \mathrm{H.c.}\,. \label{ksnl}
\end{eqnarray}
where we take $\kappa_{K^*N\Lambda}=2.43$ as used in
Refs.~\cite{stoks99,oh06}.

Second, for the u-channel nucleon exchange, the effective
Lagrangians for the $\eta NN$ and $KN\Lambda$ couplings are taken
as~\cite{tsushima97}
\begin{eqnarray}
{\cal L}_{\eta NN} &=& g_{\eta NN}{\bar N}\gamma_5 N\eta, \\
{\cal L}_{KN\Lambda} &=& g_{KN\Lambda}\bar{N}\gamma_5\Lambda
K+\mathrm{H.c.}.
\end{eqnarray}

Third, for the intermediate $\Lambda(1670)$
($\Lambda^*_{\frac{1}{2}^-}$) resonance contribution in the s-channel,
the effective Lagrangians for the $KN \Lambda^*_{\frac{1}{2}^-}$ and
$\eta\Lambda\Lambda^*_{\frac{1}{2}^-}$ vertices are~\cite{zouprc03}
\begin{eqnarray}
{\cal L}_{KN\Lambda^*_{\frac{1}{2}^-}} &=& g_{KN\Lambda^*_{\frac{1}{2}^-}}\bar{K}\bar\Lambda^*_{\frac{1}{2}^-} N+\mathrm{H.c.},\\
{\cal L}_{\eta\Lambda\Lambda^*_{\frac{1}{2}^-}} &=&
g_{\eta\Lambda\Lambda^*_{\frac{1}{2}^-}}{\bar\Lambda^*_{\frac{1}{2}^-}}{\Lambda}\eta+\mathrm{H.c.}.
\end{eqnarray}

Fourth, for the intermediate $\Lambda^*$ resonance ($D_{03}$
state) in the s-channel with spin-parity $J^P={3\over 2}^-$, the
effective Lagrangians are~\cite{xiezouliu}
\begin{eqnarray}
{\cal L}_{KN\Lambda^*_{\frac{3}{2}^-}} &=&
{f_{KN\Lambda^*_{\frac{3}{2}^-}}\over
m_K}\partial_\mu\bar{K}\bar\Lambda^{*\mu}_{\frac{3}{2}^-} \gamma_5
N+\mathrm{H.c.},\\
{\cal L}_{\eta\Lambda\Lambda^*_{\frac{3}{2}^-}} &=& {
f_{\eta\Lambda\Lambda^*_{\frac{3}{2}^-}}\over
m_\eta}\partial_\mu\eta{\bar\Lambda}^{*\mu}_{\frac{3}{2}^-}\gamma_5{\Lambda}+\mathrm{H.c.}.
\end{eqnarray}

To take into account the internal structure of hadrons and possible off-shell effects,
we introduce form factors in the amplitudes. In the present work, we
adopt the following form factors~\cite{wufq,xie,Mosel,feuster}
\begin{equation}\label{FB}
F_B(q_{ex}^2,M_{ex})={\Lambda^4\over
\Lambda^4+(q_{ex}^2-M_{ex}^2)^2}\, ,
\end{equation}
for the s and u channels and
\begin{equation}\label{FB2}
F_B(q_{ex}^2,M_{ex})=\left
(\frac{\Lambda^2-M_{ex}^2}{\Lambda^2-q_{ex}^2}\right )^2\, ,
\end{equation}
for the t channel, where $q_{ex}$ and $M_{ex}$ are the four-momenta and
the mass of the exchanged hadron, respectively. For the cutoff
parameters, we adopt $\Lambda=2.0$ GeV for the s channel and $\Lambda=1.5$
GeV for the t and u channels.

For the propagators with four-momenta $q_{ex}$, we take\cite{shyam}
\begin{equation}
G^{\mu\nu}_{K^*}(q_{ex})={-g^{\mu\nu}+q_{ex}^\mu q_{ex}^\nu/
m_{K^*}^2\over q_{ex}^2-m_{K^*}^2},
\end{equation}
for $K^*$ meson exchange, where $\mu$ and $\nu$ are polarization
indices of vector meson $K^*$.

For the proton propagator, we take
\begin{equation}
G_{N}(q_{ex}) = \frac{{\not \! q_{ex}}+M_N}{q_{ex}^2-M^2_N}.
\end{equation}

For the $\Lambda$ resonance with spin 1/2 in the s channel, we take
\begin{equation}
G_{\Lambda^*_{\frac{1}{2}}}(q_{ex}) = \frac{{\not \!
q_{ex}}+M}{q_{ex}^2-M^2+iM\Gamma},
\end{equation}
while for the $\Lambda$ resonance with spin 3/2 in the s channel, we
take the propagator as,
\begin{eqnarray}
&& G^{\mu\nu}_{\Lambda^*_{\frac{3}{2}}} (q_{ex}) =  {{\not\!
q_{ex}}+M\over q_{ex}^2-M^2+iM\Gamma} \times \nonumber \\
&& \Big(-g^{\mu\nu}+{ \gamma^\mu\gamma^\nu\over 3}+{\gamma^\mu
q_{ex}^\nu -\gamma^\nu q_{ex}^\mu\over 3M}+{2q_{ex}^\mu
q_{ex}^\nu\over 3M^2}\Big),
\end{eqnarray}
where $M$ and $\Gamma$ are the mass and width of the corresponding
intermediate state, respectively.

The differential cross section for $K^-p \to \eta \Lambda$ with the
invariant mass squared $s=(p+k)^2$ (where $k$ and $p$ are the four-momenta of the
$K^-$ and the proton) in the center-of-mass (c.m.) frame can be
expressed as
\begin{equation}
{d\sigma_{\eta\Lambda}\over d\Omega}={d\sigma_{\eta\Lambda}\over
2\pi d\cos\theta}={1\over 64\pi^2 s}{ |\vec{q}| \over
|\vec{k}|}\bar{|{\cal M}|}^2, \label{dcs}
\end{equation}
where $\theta$ denotes the angle of the outgoing $\eta$ relative to
beam direction in the c.m. frame. In the above equation, $|\vec{k}|$
and $|\vec{q}|$ denote the magnitude of the three-momenta of initial and
final state in the c.m. frame, respectively.

With the effective Lagrangian densities given above, the averaged
scattering amplitude squared $\bar{|{\cal M}|}^2$, introduced in
Eq.~(\ref{dcs}), can be expressed as
\begin{eqnarray}
\bar{|{\cal M}|}^2&=&{1\over 2}\sum_{r_1,r_2}{|\cal M|}^2\nonumber\\
&=&{1\over 2}\mathrm{Tr}\big[({\not \! p'}+m_{\Lambda}){\cal
A}({\not \! p}+m_N)\gamma^0{\cal A^+}\gamma^0\big],
\end{eqnarray}
where $r_1$ and $r_2$ denote the polarizations of the initial proton
and the final $\Lambda$, respectively; and $p$ and $p'$ denote the
four-momenta of the proton and the $\Lambda$, respectively. $\cal A$ is part of the total
amplitude, which can be expressed as
\begin{eqnarray}
{\cal M}=\bar u_{r_2}(p')~{\cal A}~u_{r_1}(p) =\bar
u_{r_2}(p')\big(\sum_\alpha{\cal A}_\alpha e^{i\phi_\alpha}
\big)u_{r_1}(p),
\end{eqnarray}
where $\alpha$ denotes the t-channel, u-channel, and various
s-channel resonances that contribute to the total amplitude. In
phenomenological approaches, the relative phase between the
amplitudes is not fixed. In our work, they are introduced as free
parameters, i.e. $\phi_\alpha$, and we take $\phi=0$ for the
amplitude of the s-channel $\Lambda(1670)$ exchange.

The $\Lambda$ polarization in the $K^-p \to \eta\Lambda$ reaction can be
studied from the decay of $\Lambda \to \pi N$. For $\Lambda \to \pi
N$, we take the following effective Lagrangian
\begin{equation}
{\cal L}_{\Lambda\pi N}=G_F m_{\pi}^2{\bar N}(A-B\gamma_5)\Lambda,
\end{equation}
where $G_F$ denotes the Fermi coupling constant, while $A$ and $B$
are effective coupling constants, for which we take
$A=1.762-0.238i$ and $B=12.24$~\cite{gaopz2011} in our calculation.

The differential cross section for $K^-p\to\eta\Lambda\to\eta\pi N$
can be expressed as
\begin{equation}
{d\sigma_{K^-p\to\eta\Lambda\to\eta\pi N}\over d\Omega
d\Omega'}={|{\bf q}||{\bf p}'_n||\bar{\cal{M}'}|^2\over 2^{11}\pi^4
m_\Lambda^2\Gamma_\Lambda s|\bf k|},
\end{equation}
where ${\bf p}'_n$ is the three-momenta of the produced nucleon in the
$\Lambda$ rest frame, $\Gamma_\Lambda=\tau^{-1}_\Lambda$ is the
$\Lambda$ decay width, and $d\Omega'=d\cos\theta'd\phi'$ is the
sphere space of the outgoing nucleon in the $\Lambda$ rest frame.
The scattering amplitude $\cal M'$ is expressed as
\begin{eqnarray}
{\cal M}' & =& \bar u_{r_3}(p_n)G_F m_\pi^2 (A-B\gamma_5)({\not \!
p'}+m_{\Lambda}) \times \nonumber \\
&& \big(\sum_\alpha{\cal A}_\alpha e^{i\phi_\alpha} \big)u_{r_1}(p),
\end{eqnarray}
and \begin{equation} |\bar{\cal{M}'}|^2={1\over
2}\sum_{r_1,r_3}{\cal M'}{\cal M'^+},
\end{equation}
with $r_1$ and $r_3$ the polarizations of the initial proton and the
final nucleon, respectively

With the above ingredients, the $\Lambda$ polarization in
$K^-p\to\eta\Lambda\to\eta\pi N$ can be expressed as

\begin{equation}
P_\Lambda={3\over \alpha_\Lambda} \Big (\int \cos\theta' {d\sigma_{
K^-p\to\eta\Lambda\to\eta\pi N}\over d\Omega d\Omega'}d\Omega'\Big
)\Big/ {d\sigma_{\eta\Lambda}\over d\Omega}, \label{pol_def}
\end{equation}
where $\alpha_\Lambda$ is the $\Lambda$ decay asymmetry parameter
with the value of $0.65$, while $\theta'$ is the angle between the
outgoing nucleon and the vector $\bf{V}=\bf k\times \bf q$.

\section{Results and Discussion}

\begin{table*}[htbp]
\caption{The fitted parameters.}
\begin{tabular}{|c|c|c|c|c|}
\hline
\hline  Channel/exchanging particle   & Product of coupling constants &$\phi_\alpha$&Mass(MeV) &Width(MeV)\\
 \hline
s/ $\Lambda(1670)$& $0.3\pm 0.03$& $0.$& $1672.5\pm 1.0$&
$24.5\pm 2.7$
\\ \hline
s/ $\Lambda(D_{03})$& $28.2\pm 7.9$& $5.66\pm 0.47$&
$1668.5\pm 0.5$& $1.5\pm 0.5$
\\ \hline
t/ $K^*$& $-58.0\pm 7.2$& $2.64\pm 0.18$& 892.& --
\\ \hline
u/ p& $-5.3\pm 1.0$& $2.59\pm 0.43$& 938.2& --
\\ \hline\hline
\end{tabular}
\label{tab1}
\end{table*}

\begin{figure}[htbp] \vspace{0.cm}
\begin{center}
\includegraphics[scale=0.8]{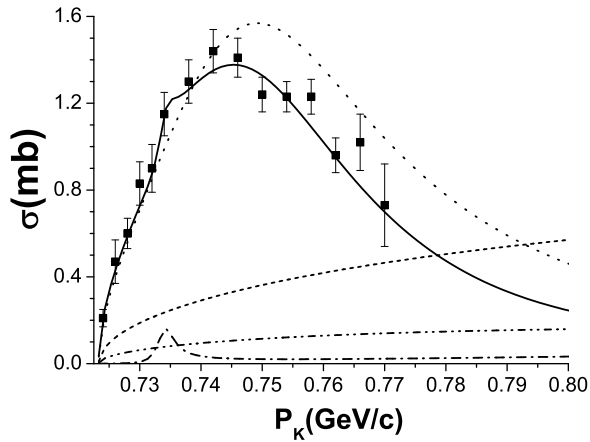}
\caption{The best fitting results for total cross sections by
considering the $\Lambda(1670)$, narrow $D_{03}$ resonance and
background contributions. The solid line represents the full
results. The contributions from the $\Lambda(1670)$, the narrow $D_{03}$
resonance, the t-channel diagram, and the u-channel diagram are shown by the
dotted, dash-dotted, dashed, and dot-dot-dashed lines, respectively. }
\label{1690}
\end{center}
\end{figure}

In the following, we present the theoretical results of a $\chi^2$ fit
to the experimental total and differential cross section
data~\cite{data} with eleven parameters ($M_{\Lambda(1670)}$,
$\Gamma_{\Lambda(1670)}$, $g_{\Lambda(1670) {\bar
K}N}g_{\Lambda(1670) \Lambda\eta}$, $M_{\Lambda(D_{03})}$,
$\Gamma_{\Lambda(D_{03})}$, $f_{\Lambda(D_{03}) {\bar
K}N}f_{\Lambda(D_{03}) \Lambda\eta}$, $g_{K^*N\Lambda}g_{K^*K\eta}$,
$g_{KN\Lambda}g_{\eta NN}$, $\phi_{\Lambda(D_{03})}$, $\phi_{K^*}$,
and $\phi_p$), which are not shown in Ref.~\cite{Liu} due to that paper's length
limit. The best fitting results for these parameters are shown in
Table~\ref{tab1}. The resultant $\chi^2/dof$ is $0.9$(where dof is the degrees of freedom).  We show in
Fig.~\ref{1690} the best fitting results for total cross sections by
considering the $\Lambda(1670)$, narrow $D_{03}$ resonance, and
background contributions. The solid line represents the full
results. The contributions from the $\Lambda(1670)$, the narrow $D_{03}$
resonance, the t-channel diagram, and the u-channel diagram are shown by the
dotted, dash-dotted, dashed, and dot-dot-dashed lines, respectively.

It was found that the $\Lambda(1670)$ resonance is needed to
interpret the steep rise of the total cross sections near threshold
and a $\Lambda(D_{03})$ state with mass $M = 1668.5 \pm 0.5$ MeV and
width $\Gamma = 1.5 \pm 0.5$ MeV is necessary to reproduce the
experimental data of the angular distributions~\cite{Liu}. Because
of its very narrow width, this $D_{03}$ resonance is obviously not
the conventional $\Lambda(1690)$ resonance
($M_{\Lambda(1690)}=1690\pm 5$ MeV, $\Gamma_{\Lambda(1690)}= 60 \pm
10$ MeV) or other $\Lambda$ states listed by the PDG.

\begin{figure}[htbp] \vspace{0.cm}
\begin{center}
\includegraphics[scale=0.8]{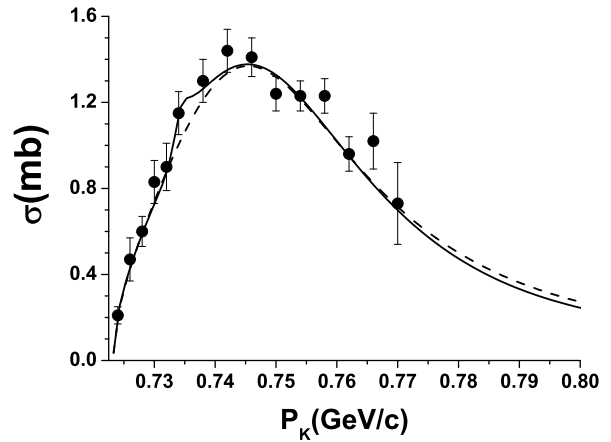}
\caption{Total cross section with(solid line)/without(dashed line)
contributions from the narrow $D_{03}$ state.}
\label{1690x}
\end{center}
\end{figure}

In Fig.~\ref{1690x} we show again our results, for comparison, for
the total cross section by including the contribution from the
narrow $D_{03}$ state with a solid line and the results obtained
without the narrow $D_{03}$ state with a dashed line. It is
clear that the small bump around $P_{K^-} = 734$ MeV can be
well reproduced if we include the contributions from the narrow
$D_{03}$ state.~\footnote{At the point $P_{K^-} = 734$ MeV, the
experimental result is $\sigma = 1.15 \pm 0.10$ mb, while our model
result, by including the narrow $D_{03}$ state, is $1.14 \pm 0.05$
mb with the error obtained from the errors of the fitted parameters
shown in Table~\ref{tab1}. We find that the agreement between our
model and the experimental result is very good. However, if we did
not consider the contribution from this narrow $D_{03}$ state, then
the theoretical result is $0.99 \pm 0.03$ mb, which gives
discrepancies of about two standard deviations with the experimental data
when the theoretical uncertainties are also taken into account.}

\begin{figure}[htbp] \vspace{0.cm}
\begin{center}
\includegraphics[scale=0.8]{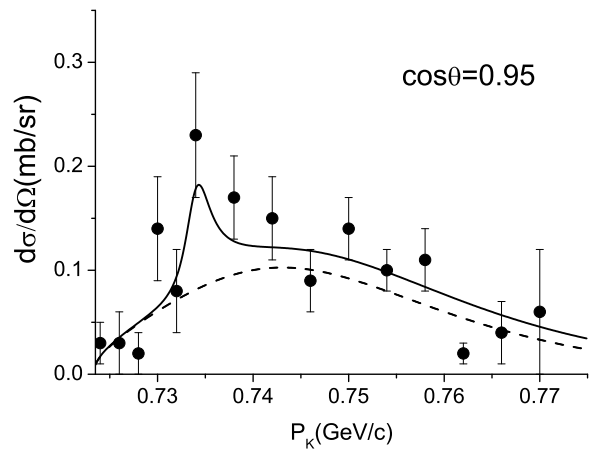}
\caption{Differential cross sections as a function of the momentum
of $K^-$ at the forward angle (cos~$\theta=0.95$) and the
corresponding theoretical results by using the best-fitted
parameters with (solid line) and without (dashed line) the narrow
resonance. } \label{forward}
\end{center}
\end{figure}

To get more clues about the role of the narrow $D_{03}$ state in the
$K^- p \to \eta \Lambda$ reaction, we calculate the differential
cross sections as a function of the momentum of the $K^-$ meson at the
forward angle (cos $\theta=0.95$). We show our results in
Fig.~\ref{forward} by comparing with the experimental data taken
form Ref.~\cite{data}. The result including the narrow
$D_{03}$ state is shown by the solid line, while the dashed line
stands for the results without including this narrow state. The bump
in the differential cross section is more clear than in the total
cross section, and this clear bump can be well reproduced by
including the contributions from the narrow $D_{03}$ state.

\subsection{The role of the $\Lambda(1690)$}

\begin{figure*}[htbp] \vspace{0.cm}
\begin{center}
\includegraphics[scale=1.2]{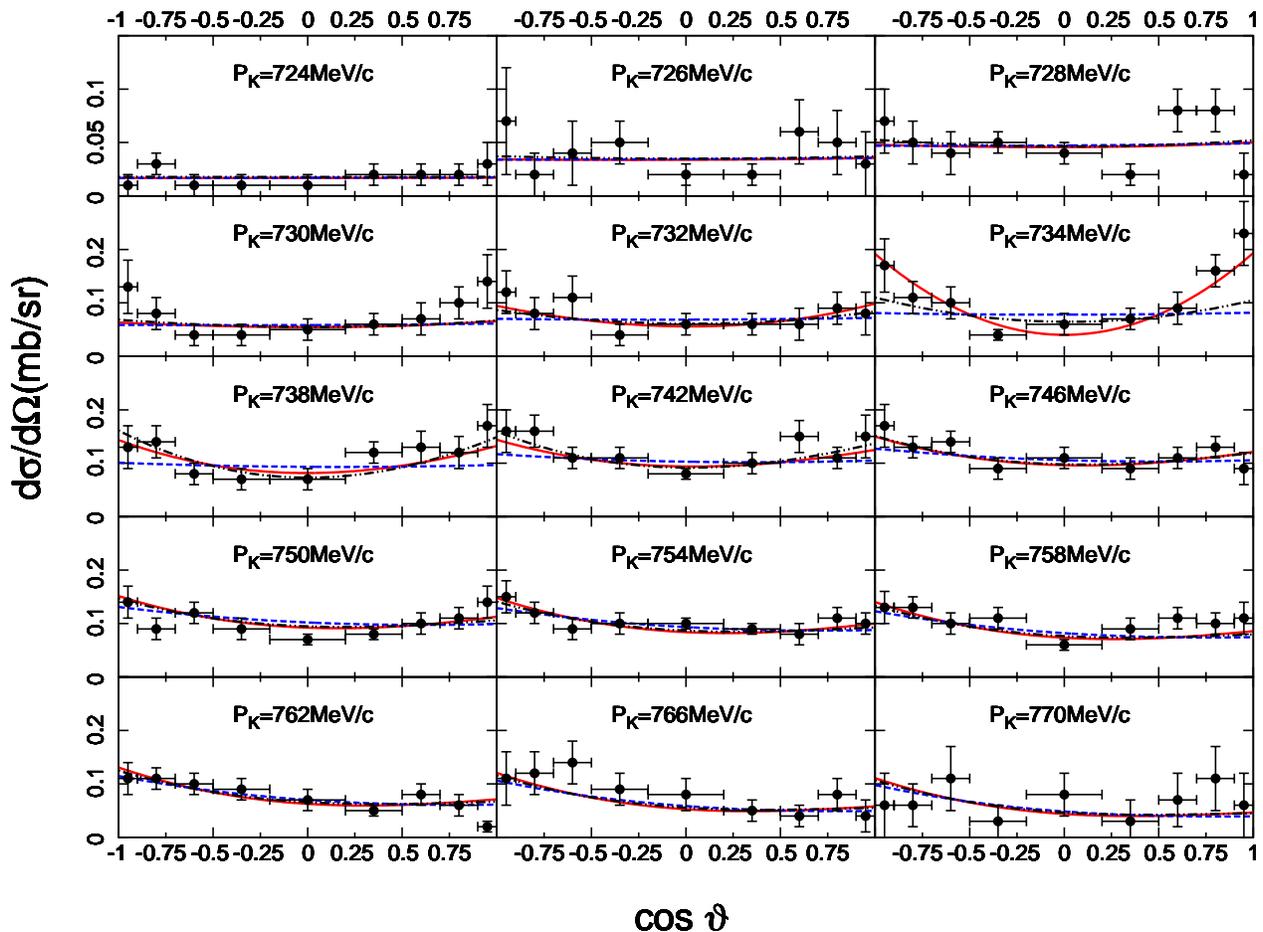}
\caption{(Color online) The fitting results for angular distributions
within Fit I (dashed line) and Fit III (dash-dot-dotted line). The
solid line represents the best fitting results taken from
Ref.~\cite{Liu} for comparison. } \label{test}
\end{center}
\end{figure*}

It is known that a well-established $D_{03}$ resonance
[$\Lambda(1690)$] may also give contributions to this reaction, and
in Ref.~\cite{data}, the authors argued that the bowl structure may
be caused by the $\Lambda(1690)$ resonance. However, it was shown in
our previous work~\cite{Liu} that the experimental data favor a
resonance with very narrow width. The main reason for the need of a
narrow resonance is because of the bowl structures only appearing in
a very narrow energy window as we have pointed out in
Ref.~\cite{Liu}.

Regarding the uncertainties of the current experimental data, it will
be interesting to discuss how well the experimental data can be
explained by the conventional $\Lambda(1690)$ state. For this
purpose, we perform the fitting procedures with some different
strategies. First, we fix the mass and width of the $D_{03}$ state
at the central values of the conventional $\Lambda(1690)$ state as given by the
PDG~\cite{pdg2012} (Fit I), i.e. mass $M=1690$ MeV and total decay
width $\Gamma=60$ MeV. Noting that the coupling of the $\Lambda(1690)$
with the $\eta\Lambda$ is consistent with zero if we take the decay
branch ratio from the PDG, here we take it as a free parameter. In
this scheme, we have nine free parameters in total. The best fitting
results are shown by the dashed line shown in Fig.~\ref{test}. It is
easy to see that the bowl structures cannot be reproduced. The
fitting results favor a weak coupling of the $\Lambda(1690)$ with the
$\eta\Lambda$ channel. This is mainly because the bowl structures
only appear in a very narrow energy window. Since the highest c.m. energy of this set of data is around $\sqrt{s}=$1.685 GeV,
which is the lower limit of the mass of the $\Lambda(1690)$ suggested by the
PDG, if the $\Lambda(1690)$ gives significant contributions to the
angular distributions, one can expect that with increasing beam
momenta the bowl structures shown in angular distributions should
become more and more prominent. However, such an expectation is not
supported by the experimental data. Therefore, we do not think the
bowl structure is caused by the conventional $\Lambda(1690)$.

Another interesting thing is to check to what extent the
experimental data can be understood by a $D_{03}$ resonance with a
normal total decay width. We then perform another fit by fixing the
width of the $D_{03}$ state at 60 MeV (Fit II), which is the width of
the conventional $\Lambda(1690)$ resonance suggested by the PDG. In this
fit, the best fitting results favor the mass $M = 1659.5 \pm 11.7$
MeV for the $D_{03}$ resonance and a small coupling with the $\eta
\Lambda$ channel. The corresponding results for angular
distributions are not shown in Fig.~\ref{test}, because they almost
overlap with the dashed-blue lines which are obtained from Fit
I.~\footnote{This is because in both Fit I and Fit II, the $D_{03}$
contributions are highly suppressed with the small coupling with
the $\eta\Lambda$ channel, so the dominant contributions are from the
$\Lambda(1670)$ resonance and background contributions which are
similar in these two fits.} So, in this fit, the bowl structures
cannot be reproduced, either.

Furthermore, since the bowl structure shows most significantly at
$P_K=$734 MeV, we also check the dependence of the results on the
data at this single energy point. In this scheme, we take the mass
and width of the $D_{03}$ resonance as free parameters and omit the
angular distribution data at $P_K=$734 MeV in the fit (Fit III). The
best fitting results still favor a narrow resonance, i.e.,
M=1670.2$\pm$1.6 MeV and $\Gamma$=4.5$\pm$1.6 MeV. This shows that
the needs of a narrow resonance are not only from one single energy
measurement but also from the pattern of angular distributions in a
wide energy region. Obviously, the description of angular
distributions in this fit (shown by the dash-dot-dotted line in
Fig.~\ref{test}) is better than for Fit I, and also for Fit II. However,
the angular distribution data at $P_K=$734 MeV is not totally
reproduced in this fit. We also show again, in Fig.~\ref{test}  by the
solid-red line, the best fitting results taken from Ref.~\cite{Liu}
for comparison.

\subsection{$\Lambda$ polarization}

\begin{figure}[htbp] \vspace{0.cm}
\begin{center}
\includegraphics[scale=0.3]{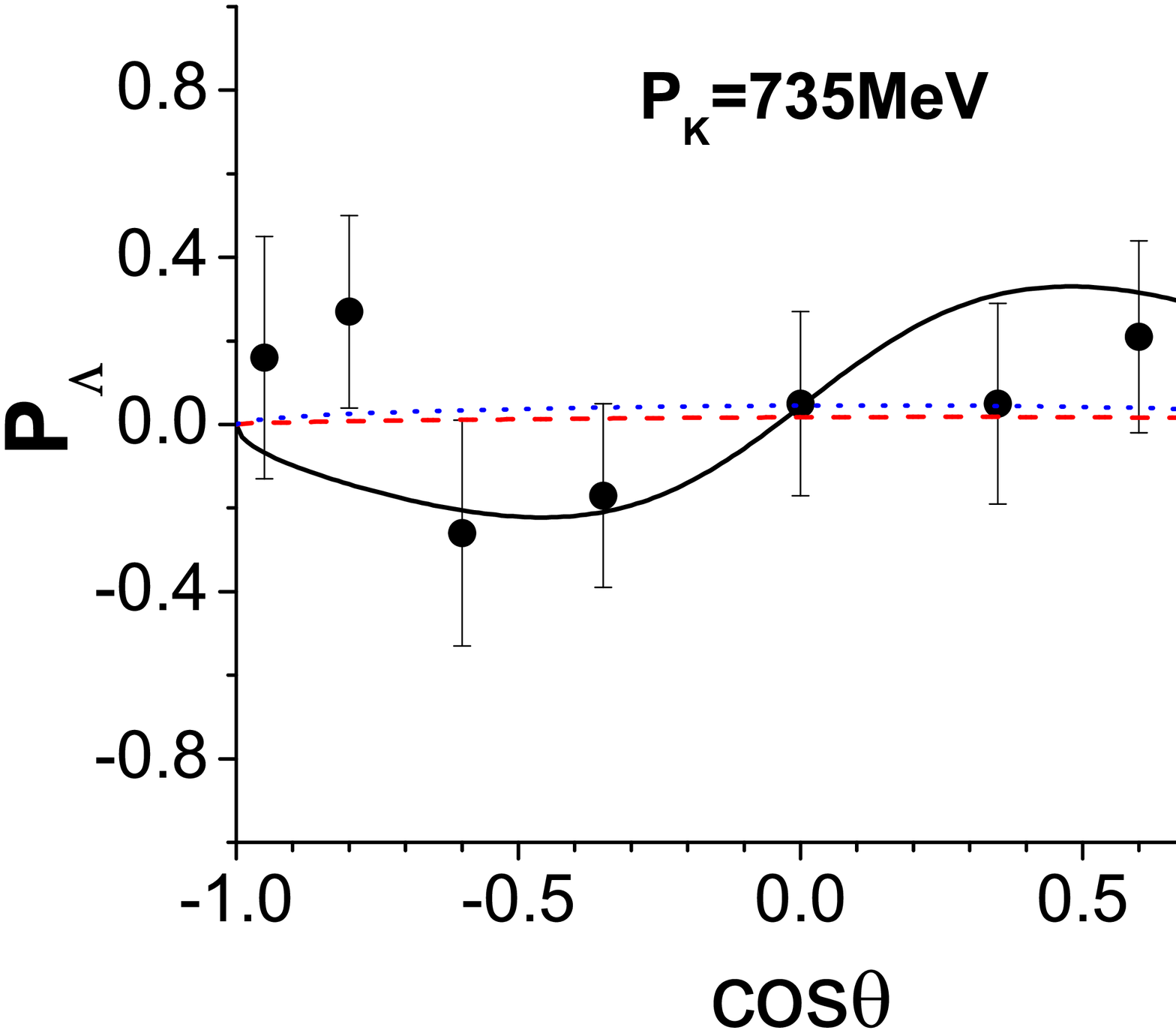}
\includegraphics[scale=0.3]{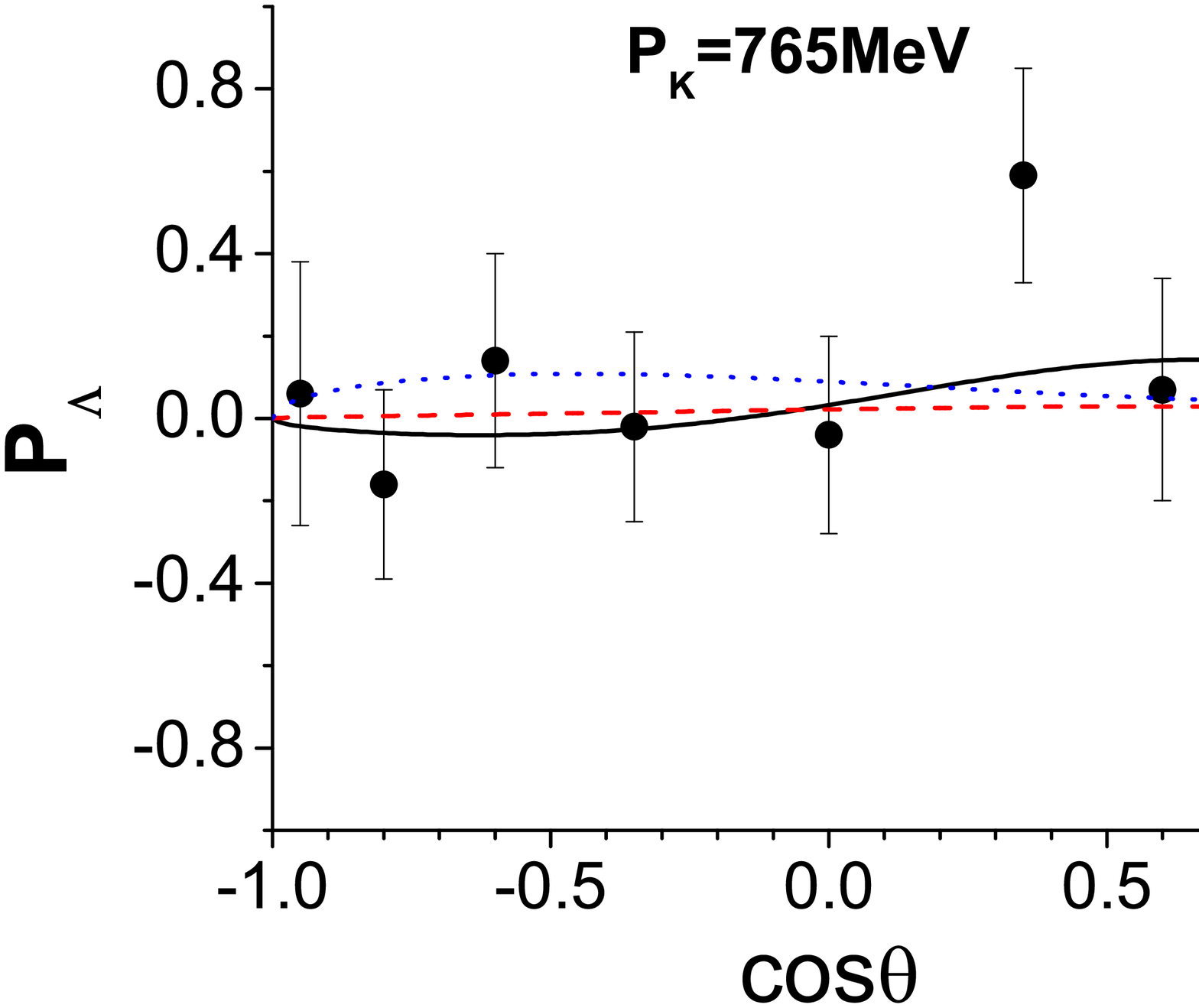}
\caption{(Color online) The predictions for $\Lambda$ polarization
with (solid line) and without (dashed line) the narrow $D_{03}$
resonance. The dotted line represents the corresponding results by
using the normal baryon resonance total decay width $\Gamma=100$ MeV
for the $D_{03}$ resonance.} \label{pol}
\end{center}
\end{figure}

It is believed that the polarization data can put more constraints
on the theoretical model and offer additional physical observables
to test the models. In Ref.~\cite{Liu}, we did not include the
$\Lambda$ polarization data in our fitting because the quality of
these data is rather poor. However, it should be meaningful to show
the predictions of our fitting results for these observables.
Following the formula and ingredients given at the end of Sec. II, we calculate the $\Lambda$ polarization in the $K^-p\to
\eta\Lambda \to \eta\pi N$ reaction at $P_{K^-} = 735$ and 765
MeV. It should be noted that all the parameters used here are taken
from Ref.~\cite{Liu} and there is no free parameter in the present
calculation. The corresponding results are shown by the solid line
in Fig.~\ref{pol}. The experimental data are taken from
Ref.~\cite{data}. By looking at Fig.~\ref{pol}, one can find
that the predictions of our model can fairly well describe the
experimental data especially at $P_{K^-}=735$ MeV, where the narrow
$D_{03}$ resonance gives a significant contribution.

It will also be interesting to check the predictions without the
narrow $D_{03}$ resonance. By using the parameters shown in Table 1 of
Ref.~\cite{Liu}, i.e., the best fitting results without the narrow
$D_{03}$ state, the corresponding $\Lambda$ polarizations are
calculated and shown by the dashed line in Fig.~\ref{pol}. It can be
found that without the narrow $D_{03}$ resonance the model
predictions can also reproduce the data within the large error bars
of experimental data, although the trend of central values of
experimental data are described badly.

We also performed the calculations by taking a normal total decay
width $\Gamma=100$ MeV for the $D_{03}$ resonance and leaving other
parameters unchanged. This calculation is meaningful because the
$\Lambda$ polarization is the ratio of two differential cross
sections [see Eq.~(\ref{pol_def})]. The corresponding results are
shown by the dotted line in Fig.~\ref{pol}.

\begin{figure}[htbp] \vspace{0.cm}
\includegraphics[scale=0.3]{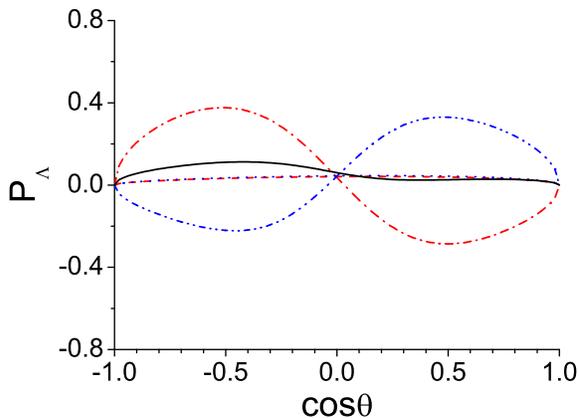}
\caption{(Color online) The results for $\Lambda$ polarization at
$P_K$=733 MeV (dash-dotted line), $P_K$=734 MeV (solid line) and
$P_K$=735 MeV (dash-dot-dotted line). The dashed and dotted lines
represent the corresponding results by omitting the $D_{03}$ state
contribution at $P_K$=733 MeV and $P_K$=735 MeV respectively.}
\label{pol_test}
\end{figure}

Besides the predictions of $\Lambda$ polarization at some discrete
energy points, it will also be interesting to investigate the energy
dependence of $\Lambda$ polarization around the momentum $P_K=734$
MeV, where the narrow $D_{03}$ resonance gives significant
contributions. In Fig.~\ref{pol_test}, we show the calculated
results for $\Lambda$ polarization at $P_K$=733, 734, and 735 MeV,
respectively, where the predictions by omitting the contribution
from the $D_{03}$ resonance at $P_K$=733 and 735 MeV are also shown for
comparison. The main finding is that the trend of $\Lambda$
polarization versus cos~$\theta$ is very sensitive to the beam
momentum in the energy range around the peak of the $D_{03}$ resonance.
After omitting the contributions of the $D_{03}$ resonance, the
results at $P_K$=733 and 735 MeV are almost flat and overlap
each other. This indicates that such energy dependence is caused by
the narrow $D_{03}$ resonance. So experimental analysis on the
energy dependence of $\Lambda$ polarization may offer further
evidence on the existence of this narrow resonance.

\section{summary}

In this work, we present some detailed analyses on the reaction
mechanism of $K^- p \rightarrow \eta \Lambda$ near threshold and
study the $\Lambda$ polarization in the $K^- p \to \eta \Lambda$
reaction based on our previous work~\cite{Liu}. It is found that
current data indeed favor a $D_{03}$ resonance with very narrow
width, which is not the conventional $\Lambda(1690)$. And the
existence of this narrow $D_{03}$ state is also compatible with the
current $\Lambda$ polarization data. We also study the role of the
conventional $\Lambda(1690)$ resonance in this reaction, however,
the current experimental data cannot be reproduced by including the
conventional $\Lambda(1690)$ resonance.

In the present work, we find that the $\Lambda$ polarization are
strongly energy dependent around $P_{K^-}$=734 MeV with inclusion
the narrow $D_{03}$ state . Thus more accurate polarization data
around $P_{K^-}=734$ MeV can be used to verify the existence of this
$D_{03}$ resonance. We suggest our colleagues
remeasure both the differential cross section and $\Lambda$
polarization data, which should be helpful to clarify the existence
of this $D_{03}$ resonance.

\section*{Acknowledgments}

We would like to thank Xian-Hui Zhong, Pu-Ze Gao, and Jia-Jun Wu for
useful discussions. We also want to thank Prof. Bing-Song Zou for
his valuable suggestions and support during our stay at the
Institute of High Energy Physics. This work is supported by the
National Natural Science Foundation of China under Grants No. 10905046
and No. 11105126. B. C. Liu is also supported by the Fundamental
Research Funds for the Central Universities.


\begin{thebibliography}{99}

\bibitem{pdg2012} J.~Beringer \emph{et al.}, Phys. Rev. {\bf D86},
010001 (2012).
\bibitem{data} A. Starostin, \emph{et al.}, Phys. Rev. {\bf C64},
055205 (2001).
\bibitem{Klempt2010} E. Klempt and J.-M. Richard, Rev. Mod. Phys. {\bf 82}, 1095 (2010).

\bibitem{gaopz2011} P. Z. Gao, B. S. Zou and A. Sibirtsev, Nucl. Phys.
{\bf A867} (2011) 41.
\bibitem{lam1670prl} D. M. Manley, \etal, Phys.\ Rev.\ Lett. \textbf{88}, 012002 (2001).
\bibitem{oset1} E. Oset, A. Ramos, C. Bennhold, Phys.\ Lett.\ B \textbf{527},
99 (2002).
\bibitem{oset2} C. Garcia-Recio, J. Nieves, E. Ruiz Arriola and M. J. Vicente
Vacas, Phys. Rev. {\bf D67}, 076009 (2003).
%
\bibitem{zhongprc79} Xian-Hui Zhong, and Qiang Zhao, Phys. Rev. \textbf{C79}, 045202 (2009).
%
\bibitem{Liu} Bo-Chao Liu and Ju-Jun Xie, Phys. Rev. \textbf{C85}, 038201 (2012).
%
\bibitem {stoks99} V. G. J. Stoks and Th. A. Rijken,
Phys. Rev. {\bf C59}, 3009 (1999).
%
\bibitem{oh06} Y. Oh and H. Kim,
Phys. Rev. {\bf C73}, 065202 (2006).
\bibitem{tsushima97} K. Tsushima, A. Sibirtsev and A. W. Thomas, Phys. Lett. {\bf B390},29 (1997).
\bibitem{zouprc03} B.~S.~Zou and F.~Hussain, Phys.\ Rev.\ C \textbf{67}, 015204 (2003).
\bibitem{xiezouliu} Ju-Jun Xie, Bing-Song Zou, and Bo-Chao Liu, Chin.
Phys. Lett. {\bf 22}, 2215 (2005).
%
\bibitem{wufq} F. Q. Wu, B. S. Zou, L.Li and D. V. Bugg, Nucl. Phys.
{\bf A735} (2004) 111.
\bibitem{xie} J. J. Xie, B. S. Zou and H. Q. Chiang, Phys. Rev. {\bf
C77},015206(2008).
\bibitem{Mosel} G. Penner and U. Mosel, Phys. Rev. C {\bf 66}, 055211
(2002); ibid. C {\bf 66}, 055212 (2002);\\ V. Shklyar, H. Lenske and
U. Mosel, Phys. Rev. C {\bf 72}, 015210 (2005).
\bibitem{feuster} T. Feuster and U. Mosel, Phys. Rev. C
\textbf{58}, 457 (1998);\\
T. Feuster and U. Mosel, Phys. Rev. C \textbf{59}, 460 (1999).
\bibitem{shyam} R. Shyam, Phys. Rev. C \textbf{60}, 055213 (1999).
\end{thebibliography}
\end{document}